\documentclass[preprint,12pt]{elsarticle}
\usepackage{amsmath}
\usepackage{amssymb}
\usepackage[pdftex,colorlinks]{hyperref}
\usepackage{multirow}

\biboptions{compress}

\begin{document}

\begin{frontmatter}

\title{Quantum fully homomorphic encryption scheme based on quantum fault-tolerant construction}

%% use optional labels to link authors explicitly to addresses:
\author[label1]{Min Liang}\ead{liangmin07@mails.ucas.ac.cn}
\author[label2]{Li Yang\corref{1}}\ead{yangli@iie.ac.cn}
\cortext[1]{Corresponding author.}
\address[label1]{Data Communication Science and Technology Research Institute, Beijing 100191, China}
\address[label2]{State Key Laboratory of Information Security, Institute of Information Engineering, Chinese Academy of Sciences,
Beijing 100093, China}

\begin{abstract}
Fully homomorphic encryption is a kind of encryption scheme, which enables arbitrary computation on encrypted data without accessing the data. We present the quantum version of fully homomorphic encryption scheme, which is constructed based on quantum fault-tolerant construction. Two schemes are constructed. The first is a symmetric scheme, and the secret key is the quantum CSS code. In the scheme, when Server performs quantum computation on the encrypted plaintext, some ancillary quantum states should be provided by Client. The second is an asymmetric scheme, which contains the periodical interaction between Client and Server.
\end{abstract}

\begin{keyword}
Quantum cryptography \sep fully homomorphic encryption \sep quantum CSS codes \sep fault-tolerant quantum computation
%% keywords here, in the form: keyword \sep keyword
%% MSC codes here, in the form: \MSC code \sep code
%% or \MSC[2008] code \sep code (2000 is the default)
\end{keyword}

\end{frontmatter}

%%
%% Start line numbering here if you want
%%
% \linenumbers
\section{Introduction}
Fully homomorphic encryption (FHE) is a kind of encryption scheme with the homomorphic property. It enables arbitrary computation on encrypted plaintext, without decrypting them. FHE is fundamental in the theory and application of cryptology, such as secure multiparty computation, zero-knowledge proof, search encrypted data, outsourcing computing, copyright protection. For many years, the researchers yearn for a construction of FHE scheme. Since Gentry \cite{gentry2009} sucessfully devised a FHE scheme, FHE becomes an important research direction in cryptology.

Quantum fully homomorphic encryption (QFHE) \cite{liang2013} represents a quantum version of FHE. Generally, QFHE is a kind of FHE scheme which is realized through quantum computing technology. Similar with the classical FHE, a QFHE has four algorithms: key generation algorithm, encryption algorithm, decryption algorithm, evaluation algorithm. The difference between QFHE and FHE is the computing model. FHE is established under the model of polynomial-time probabilistic computation, while QFHE is defined by the model of polynomial-time quantum computation.

In 2012, Rohde et al.\cite{rohde2012} proposed a symmetric quantum homomorphic encryption scheme, which permits quantum random walk on encrypted quantum data. However, it is not fully homomorphic. Later, in Ref.\cite{liang2013}, QFHE is formalized, and a weak symmetric scheme is constructed. By the scheme, the encrypted quantum data can be directly used in the quantum computation, without decrypting them in advance. However, the scheme is different from the commonly mentioned FHE, since the secret key is necessary during the computation on encrypted data.

Recently, Yu et al.\cite{yu2014} proved that, when the perfect security is required, quantum homomorphic encryption scheme would need at least $\texttt{log}_2|S|$ qubits to store the encrypted result, where $|S|$ denotes the size of the set $S$ ($S$ is the set of permitted computation). So, the QFHE scheme with perfect security would necessarily incur exponential storage overhead. The QFHE scheme proposed in Ref.\cite{tan2014} does not provide the security of cryptographic sense. The reason is as follow: it only hides $n/\texttt{log}n$ bits ($n$ is the size of the message), then the ratio of the hidden amount to the total amount  approaches to 0 ($1/\texttt{log}n\rightarrow 0$) when $n$ approaches to infinity. Broadbent and Jeffery \cite{broadbent2014} proposed three public-key quantum homomorphic encryption schemes (not fully homomorphic), which are constructed by combining quantum one-time pad with classical FHE scheme. So, their efficiency and security are bounded by classical FHE scheme.

In modern cryptography, error-correcting code is an important tool in the design of cryptosystems, such as McEliece public-key encryption algorithm \cite{mceliece1978}. As a special lattice, error-correcting codes can also be used to construct FHE scheme. Similarly, in quantum cryptography, quantum error-correcting codes can be used to devise quantum cryptosystems, such as quantum authentication scheme \cite{barnum2002} and Fujita's public-key encryption scheme \cite{fujita2012}. In addition, we also have proposed a quantum McEliece public-key encryption scheme \cite{yang2003,yang2010,yang2015} based on error-correcting codes. In this article, we shall devise a QFHE scheme based on quantum fault-tolerant construction.

\section{Quantum fully homomorphic encryption scheme}
\subsection{Symmetric scheme}
CSS code is an important kinds of quantum error-correcting codes, and plays the fundamental role in fault-tolerant quantum computation. We devise QFHE scheme based on CSS codes, for the reason that it has good property in fault-tolerant quantum computation: after encoding quantum states through CSS codes, the quantum operation on logical qubit can be implemented by transversal quantum operations on physical qubits. For example, the fault-tolerant H gate and CNOT gate can be implemented by transversal H gates and transversal CNOT gates, respectively. The fault-tolerant S gate can be implemented by transversal $S^{-1}$ gates. Thus, when the code-length is known, the fault-tolerant H, CNOT, S gates can be carried out without any information about the code. In fault-tolerant quantum computation, the T (or $\pi/8$) gate can also be implemented transversally, however, an ancillary state is necessary, which is relative to the CSS code. See Ref.\cite{nielsen2000} for more details.

Firstly, we should determine a value of security parameter $n$. For a given value $n$, there exists $2n-1$ kinds of $[[n,1]]$ quantum CSS codes. Denote $Q_n$ as the set of all $[[n,1]]$ quantum CSS codes. The size of $Q_n$ is $|Q_n|=2^{n-1}$.

Next, we present a symmetric QFHE scheme based on the set $Q_n$. The secret key is selected from the set $Q_n$, and is used in the encryption/decryption algorithms. The scheme contains the following four algorithms.
\begin{itemize}
  \item Key generation algorithm

  Alice randomly selects a CSS code from the set $Q_n$. The CSS code is used as the secret key. Denote it as $sk$. We suggest the following two ways to randomly select the secret key.
  \begin{enumerate}
    \item Similar to the key generation algorithm of Fujita's public-key encryption scheme \cite{fujita2012}, Alice selects a CSS code of generator matrix $G$ from $Q_n$, and randomly selects a nonsingular binary matrix $S$ and a permutation matrix $P$ (The concrete parameter and form of these matrixes can be referred in \cite{fujita2012}). Then, Alice computes $\hat{G}=SGP$, which is a generator matrix of another CSS code in $Q_n$. $\hat{G}$ is used as the secret key.
    \item In the other way, it uses a family of CSS codes with parameters $u$ and $v$, denoted as $\{CSS_{u,v}(C_1,C_2),u,v\in\{0,1\}^n\}$ \cite{nielsen2000}. Firstly, Alice determines a CSS code $CSS(C_1,C_2)$, then randomly selects two random numbers $u$ and $v$. The CSS code $CSS_{u,v}(C_1,C_2)$ is used as the secret key.
  \end{enumerate}

  \item Encryption algorithm

  According to the secret key $sk$ (a CSS code), Alice performs CSS encoding on quantum plaintext $\rho$,
  \begin{equation}\sigma=Enc_{sk}(\rho),\end{equation}
  and each qubit is encoded into $n$ cipher qubits. The quantum circuit for encoding can be constructed from the standard form of the generator matrix for a stabilizer code \cite{gottesman1997}.

  \item Evaluation algorithm

  Denote $QC$ as Bob's quantum computation and $\overline{QC}$ as its fault-tolerant version. Bob does not know the secret key, which is the encoding scheme of CSS code. He performs fault-tolerant quantum computation $\overline{QC}$ on encoded quantum data $\sigma$, and obtains the encoded result $\sigma'=\overline{QC}(\sigma)$, which equals to $Enc_{sk}(QC(\rho))$. The details are as follows. The set $\{\texttt{H},\texttt{CNOT},\texttt{T}\}$ is a universal set of gates for quantum computation, so any quantum computation can be described by a quantum circuit consisting of the H,CNOT,T gates. According to the quantum circuit, Bob performs quantum computation on encrypted states in the following ways.
  \begin{enumerate}
    \item If Bob intends to perform an H (or CNOT) gate on the $i$th (or $i,j$th) logical qubits of the plaintext $\rho$, he should perform a fault-tolerant H (or CNOT) gate on the corresponding physical qubits of the encrypted states $\sigma$. The fault-tolerant H (or CNOT) has $n$ transversal H (or CNOT) gates, denoted as $\overline{H}$ (or $\overline{CNOT}$).
        \begin{eqnarray}
        \overline{H}(Enc_{sk}(\rho))=Enc_{sk}(\texttt{H}(\rho)),\\
        \overline{CNOT}(Enc_{sk}(\rho))=Enc_{sk}(\texttt{CNOT}(\rho)).
        \end{eqnarray}
    \item If Bob intends to perform a T gate on the $i$th logical qubit of the plaintext $\rho$, he needs an ancillary state $|\Theta\rangle$ which depends on the secret key (or the CSS code) $sk$. Because $sk$ is private for Bob, the ancillary state must be provided by Alice. Alice can produce the ancillary state by encrypting a qubit $(|0\rangle+exp(i\pi/4)|1\rangle)/\sqrt{2}$:
        \begin{equation}
        |\Theta\rangle=Enc_{sk}\left(\frac{|0\rangle+exp(i\pi/4)|1\rangle}{\sqrt{2}}\right).
        \end{equation}
        To avoid interactive process between Alice and Bob, the ancillary state $|\Theta\rangle$ should be prepared when Alice performs the encryption algorithm, and be sent to Bob together with the encoded state $\sigma$. According to the implementation of fault-tolerant $T$ gate \cite{nielsen2000}, by using the ancillary state, the $T$ gate on the logical qubit can also be implemented by $n$ transversal CNOT gates and $n$ transversal $S^{-1}X$ gates.
  \end{enumerate}

  \item Decryption algorithm

  Denote $\sigma'$ as the obtained result after finishing the Evaluation algorithm on the encrypted state $\sigma$. According to the secret key $sk$ (a CSS code), Alice performs CSS decoding on quantum ciphertext $\sigma'$
  \begin{equation}\rho'=Dec_{sk}(\sigma'),\end{equation}
  then every $n$ cipher qubits are decoded into single plain qubit. The quantum circuit for decoding can be constructed from the standard form of the generator matrix for a stabilizer code \cite{gottesman1997}.

\end{itemize}

Until now, we have completely described a symmetric QFHE scheme. Its correctness (or $\rho'\equiv QC(\rho)$) can be deduced from fault-tolerant quantum computation
\begin{equation}\overline{QC}(Enc_{sk}(\rho))=Enc_{sk}(QC(\rho)),\end{equation}
where $\overline{QC}$ represents the quantum computation on encoded qubits or physical qubits, and $QC$ represents the quantum computation on logical qubits.

It is worth to notice that, the ancillary state $|\Theta\rangle$ depends on the CSS code, which is Alice's secret key. Thus, the ancillary state $|\Theta\rangle$ should be provided by Alice, and their amount depends on the number of $T$ gates in Bob's quantum computation $QC$. To avoid interactive computing, Alice should provide enough ancillary states when she sends the encrypted state $\sigma$ to Bob.

Another thing that should be noticed is, our QFHE scheme is constructed based on quantum fault-tolerant construction, however, it is unnecessary to periodically perform error-correction precedure.

It can be known from our construction that, the security is independent of any computational hard problems. Finally, we analyze the security of the QFHE scheme.

Alice's secret key $sk$ is a quantum CSS code in $Q_n$, the attacker does not know which CSS code is selected by Alice. There may be two possible ways to attack the scheme: (1)Decrypt the cipher state through CSS decoding, (2)Decrypt the ancillary state $|\Theta\rangle$ through CSS decoding. In the first case, the attacker knows neither the plain state nor the CSS code $sk$. Thus, even if he guesses the secret key $sk$ correctly, he cannot judge whether he has attacked successfully. In the second case, the attacker can decode the ancillary state, and compare the decoded state with $(|0\rangle+exp(i\pi/4)|1\rangle)/\sqrt{2}$, then he can know whether he succeed. So, the focus of the security is the total amount of $[[n,1]]$ quantum CSS codes. Because the total number of $[[n,1]]$ quantum CSS codes is $2^{n-1}$, when n is large enough, the attack will succeed with an exponentially small probability. Because the ancillary state contains some information about the CSS code $sk$, the attacker can determine $sk$ from sufficiently more ancillary states. In the practice, Alice should predict how many $T$ gates would be in the quantum circuit $QC$, and select a large enough security parameter $n$. Then the attacker cannot obtain enough information about $sk$ by measuring these ancillary states.

The theoretical analysis of the security about the QFHE scheme will be given in the full paper.

\subsection{Asymmetric scheme}
We have proposed a symmetric QFHE scheme. How to construct an asymmetric scheme? In this section, we try to solve the problem based on quantum CSS codes. Similar to Fujita's public-key cryptosystems \cite{fujita2012}, random errors are added to the encoded quantum state while encrypting the state. However, this would cause a problem: when Bob performs nonlocal quantum computation on the cipher state, the random errors would propagate between different logical qubits (a logical qubit is a block of physical qubits). The errors in single logical qubit maybe increase and finally cause the failure of decryption. Similar to the periodically error-correction in fault-tolerant quantum computation, we can solve the problem by periodical interaction: when Bob finishes a period of quantum computation on encrypted data, he sends the outcome to Alice; Alice decrypts it and encrypts it again, then she sends the new cipher state to Bob; Bob begins the next period of quantum computation.
\begin{itemize}
  \item Key generation algorithm

  Alice determines the private key $(S,G,P)$ (It is the same as Ref.\cite{fujita2012}), and computes $\hat{G}=SGP$. $\hat{G}$ can determine a CSS code. The CSS code can correct up to $t$ errors. Then the public key is $(\hat{G}, ct)$, where $c\in(0,1)$ is a constant number.

  \item Encryption algorithm

  According to the public key $\hat{G}$, Alice encodes the quantum plaintext $\rho$, and randomly adds $ct$ Pauli errors (denoted as $P_{ct}$) on the encoded state,
  \begin{equation}\rho'=P_{ct}(Encode_{\hat{G}}(\rho)).\end{equation}
  Then she sends the cipher state $\rho'$ to Bob.

  \item Evaluation algorithm

  Assume Bob's quantum computation has been represented as a quantum circuit $QC$ consisting of H, T, CNOT gates. After receiving the cipher state $\rho'$ from Alice, Bob begins to perform the quantum gates in his circuit according to the public key $\hat{G}$. For the different gates, he performs in the following different ways.
  \begin{enumerate}
    \item For the H gate, Bob performs the fault-tolerant Hadamard gate $\overline{H}$ according to the CSS code $\hat{G}$. $\overline{H}$ is $n$ transversal H gates.
    \begin{equation}\overline{H}P_{ct}(Encode_{\hat{G}}(\rho))=P'_{ct}\overline{H}(Encode_{\hat{G}}(\rho))=P'_{ct}(Encode_{\hat{G}}(H(\rho))),\end{equation}
    where $P'_{ct}$ denotes another $ct$ Pauli errors. The positions of the $ct$ Pauli errors are the same as $P_{ct}$.
    \item For the T gate, Bob prepares an ancillary state $|\Theta\rangle$ according to the CSS code $\hat{G}$. Then he performs the fault-tolerant T gate.
    \item For the CNOT gate, when Bob performs a fault-tolerant CNOT gate, the Pauli errors will increase. When the total errors increase to $t$, Bob returns the outcome to Alice. According to the private key $(S,G,P)$, Alice decrypts it. Then she encrypts it again, and sends to Bob. Bob continues the unfinished quantum computation.
  \end{enumerate}
  Performing the above interactive procedure until Bob finishes the whole quantum circuit.

  \item Decryption algorithm

  According to the private key $(S,G,P)$, Alice can decrypt the cipher state. See Ref.\cite{fujita2012} for the details.

\end{itemize}

This asymmetric scheme is computationally secure. Its formal description and strict security analysis will be given in the full paper.

\section{Discussions}
In the asymmetric QFHE scheme, random Pauli errors are introduced to realize the encryption, and the interaction between Alice and Bob is introduced to control the increase of the errors. To avoid the interactive process, one possible solution is the quantum bootstrapping, which is similar to the bootstrapping in classical FHE. The quantum bootstrapping is still under consideration now.

The encoder of any quantum CSS code cannot commute with arbitrary quantum operator. From the fault-tolerant quantum computation, we know the commutative rules between the H (or CNOT) gate and any CSS encoder $Enc_{css}$.
\begin{eqnarray}
\overline{H}(Enc_{css}(\rho)) &=& Enc_{css}(H(\rho)),\forall \rho,\\
\overline{CNOT}(Enc_{css}(\rho)) &=& Enc_{css}(CNOT(\rho)),\forall \rho.
\end{eqnarray}
However, in the universal set $\{H,CNOT,T\}$, only the T gate does not satisfy the above relation. So, an ancillary state is employed for the T gate in the symmetric scheme. While in the asymmetric scheme, the interactive process between Alice and Bob is used. We guess that, in the construction of QFHE scheme, it is necessary that Alice provides ancillary computation for Bob.

\section{Conclusions}
Based on quantum fault-tolerant construction, this paper presents two QFHE schemes. In the symmetric scheme, the CSS code is used as the secret key for encryption and decryption. Quantum fault-tolerant construction is adopted during the computing on encrypted states, however there has not any periodical error-correction and no interaction is needed. The security of the scheme does not depend on any computational hard problems. In the asymmetric scheme, interaction between Alice and Bob is necessary, and the security relies on NP-complete problem.

\section*{Acknowledgement}
This work was supported by the National Natural Science Foundation of China under Grant No.61173157.


\begin{thebibliography}{00}
%\softraggedright
\itemsep=-1pt plus.2pt minus.2pt  %% sets the vertical space between items
%\small
\bibitem{gentry2009}
Gentry, C.: A fully homomorphic encryption scheme. PhD thesis, Stanford University (2009)

\bibitem{liang2013}
Liang, M.: Symmetric quantum fully homomorphic encryption with perfect security. Quantum Inf. Process. {\bf 12}, 3675-3687 (2013)

\bibitem{rohde2012}
Rohde, P. P., Fitzsimons, J. F., Gilchrist, A.: Quantum Walks with Encrypted Data. Phys. Rev. Lett. {\bf 109}(15), 150501 (2012)

\bibitem{yu2014}
Yu, L., Perez-Delgado, C. A., et al.(2014). Limitations on information theoretically secure quantum homomorphic encryption. Physical Review A, 2014.

\bibitem{tan2014}
Tan, S. H., Joshua, A. K., et al.(2014). A quantum approach to fully homomorphic encryption. arXiv:1411.5254v2.

\bibitem{broadbent2014}
Broadbent, A., Jeffery, S.: Quantum homomorphic encryption for circuits of low T-gate complexity. arXiv:1412.8766.

\bibitem{mceliece1978}
McEliece, R.: A public-key cryptosystem based on algebraic coding theory. DSN progress report 42(44):114-116 (1978)

\bibitem{barnum2002}
Barnum, H., et al.(2002). Authentication of quantum messages. The 34th Annual IEEE Symposium on Foundations of Computer Science, Compendex: 449-458.

\bibitem{fujita2012}
Fujita, H.: Quantum McEliece public-key cryptosystem, Quantum information \& computation 12(3\&4):181-202 (2012)

\bibitem{yang2003}
Yang, L.: Quantum public-key cryptosystem based on classical NP-complete problem. Arxiv preprint quant-ph/0310076. (2003)

\bibitem{yang2010}
Yang, L., et al.: Quantum public-key cryptosystems based on induced trapdoor one-way transformations. arXiv:1012.5249. (2010)

\bibitem{yang2015}
Yang, L., Liang, M.: Quantum McEliece public-key encryption scheme. arXiv:1501.04895. (2015)

\bibitem{nielsen2000}
Nielsen, M., Chuang, I.: Quantum computation and quantum information. Cambridge University Press, Cambridge (2000)

\bibitem{gottesman1997}
Gottesman, D.: Stabilizer codes and quantum error correction. PhD thesis, California Institute of Technology (1997)
\end{thebibliography}
\end{document}